 \documentclass[preprint,authoryear,11pt]{elsarticle}

 \usepackage{graphics}
\usepackage{amssymb}
\usepackage{amsthm}
\usepackage{amsmath}

\begin{document}
\newtheorem{remark}{\textbf{Remark}}
\newtheorem{theorem}{\textbf{Theorem}}
\newtheorem{definition}{\textbf{Definition}}
\newtheorem{proposition}{\textbf{Proposition}}
\newtheorem{corollary}{\textbf{Corollary}}

\begin{frontmatter}



 \title{Active margin system for margin loans using cash and stock as collateral and its application in Chinese market}

%\fntext[label3]{Department of Statistics and Actuarial Science, Chongqing
%University, Chongqing, 400030  P.R. CHINA;}


\author{Guanghui Huang\corref{cor1}}
\ead{hgh@cqu.edu.cn}

\author{Weiqing Gu}

\author{Wenting Xing}

\author{Hongyu Li}

\cortext[cor1]{Corresponding author.}
\address{College of Mathematics and Statistics, Chongqing University,
  Chongqing 401331, China}

\begin{abstract}
Margin system for margin loans using cash and stock as collateral is considered in this paper, which is the line of defence for brokers against risk associated with margin trading. The conditional probability of negative return is used as risk measure, and a recursive algorithm is proposed to realize this measure under a Markov chain model. Optimal margin system is chosen from those systems which satisfy the constraint of the risk measure. The resulted margin system is able to adjust actively with respect to the changes of stock prices. The margin system required by the Shanghai Stock Exchange is compared with the proposed system, where 25,200 margin loans of 126 stocks listed on the SSE are investigated. It is found that the number of margin calls under the proposed margin system is significantly less than its counterpart under the required system for the same level of risk, and the average costs of the loans are similar under the two types of margin systems.
\end{abstract}


\end{frontmatter}

% \linenumbers


\section{Introduction}

A margin loan is a loan made by a brokerage house to a client that allows the customer to buy stocks on credit. The term margin itself refers to the difference between the market value of the shares purchased and the amount borrowed from the broker.
As a margin loan gives customer more to invest, there is a potential for bigger returns. Of course, this also magnifies the potential for losses if the investments perform poorly \citep{fortune2000,fortune2001,ricke2003}. In order to restrain the level of speculation and to protect investors and brokers,
the United States Congress gave the Federal Reserve System the power to control margin requirements in 1934. The initial margin requirement was changed 22 times between 1934 and 1974, and it has been fixed at $50\%$ since 1974 \citep{fortune2000}.

\citet{galbraith} and \citet{hsieh1990} indicated that low margin requirement was one of the major contributing factors which led to the Stock Market Crash of 1929.
Shiller claimed in a {\it Wall Street Journal} article that the stock market crash of 1987 and the stock market boom in the late 1990s led to calls for the return to an active margin policy, which sets a minimum equity
position on the date of a credit-financed security transaction \citep{shiller2000,ricke2003}. On the other hand, Bartlett argued that although the adjustment of margin policy can not impact the volatility of stock market, brokers should set their own house margins to react to the changes of stock market, which can protect themselves from customer defaults \citep{shiller2000}. The purpose of this paper is to set an active margin system for margin loans traded in Chinese market.

The margin system in Chinese market is composed of initial margin requirement, maintenance margin requirement and mandatory liquidation, which is the line of defence for brokers against the risk associated with the transactions of margin loans. Pursuant to the regulations of the Shanghai Stock Exchange (SSE) and the Shenzhen Stock Exchange (SZSE), the initial and maintenance margin ratios are $50\%$ and $130\%$ respectively, which is called the required margin system in this paper.
However, \citet{argiriou2009} pointed out that although supervising authorities give recommendations or limits for margin requirements, there is no information on how these recommendations are derived. Therefore a risk measurement based on conditional probability is adopted in this paper, which is used to set margin system for Chinese market with a specified target of risk control.

The probability that broker yields a negative return after mandatory liquidation under the condition that a margin call has been issued is called the conditional probability of negative return (CPNR) in this paper. CPNR is used to measure the risk faced by brokers, which does not only take into account the possibility of investor's default, but also emphasize on the stop-loss function of mandatory liquidation \citep{huang2010}.  A simple margin loan is considered in this paper, where an amount of cash and a portion of the stock to be purchased are used as collateral. An active margin system is given for a particular margin account, such that the value of CPNR is less than or equal to the desired value of broker. The margin system deduced from the constraint of CPNR is called the deduced margin system in this paper.

126 stocks listed on the SSE are used to compare the performances of the required and the deduced margin systems, where 200 margin loans with a 30-day period are constructed for each stock. It is found that the average number of margin calls is reduced by $59.8\%$ under the deduced margin system  with respect to the required margin system, while the average costs under the two margin systems are similar. Those empirical results indicate that CPNR can be used to measure the risk associated with margin loans, and the proposed margin system is an operational method for brokers to set their own margin requirements.

The remainder of this paper is organized as follows. The margin requirements in Chinese market and the risk measure are introduced in Section 2. Section 3 is devoted to the realization of the risk measurement under a Markov chain model. An active margin system is given for the loans using cash and stock as collateral in Section 4. The empirical investigations of Chinese margin loan market are given in Section 5. The conclusions and discussions are given in Section 6.

\section{Risk measure for Chinese market}

\subsection{Margin system of Chinese market}

Pursuant to the regulations of the SSE and the SZSE, an amount of initial margin should be deposited to the broker on the date of margin trading, which can be composed of cash, the underlying stock, and other marginable securities listed by the exchanges. The ratio between the market value of the initial margin and the market value of the purchased stock is called the initial margin ratio, which must be larger than or equal to $50\%$. The purchased stock and the initial margin are kept in the account of collaterals. The ratio between the market value of the collaterals and the market value of the loan is called the maintenance margin ratio, which must be larger than or equal to $130\%$.

When the value of maintenance margin ratio is below the required level, a margin call will be issued by broker, which informs the customer to increase the margin that he has deposited or close out his position. If he does none of
these,  broker can sell the collateral to meet the margin
call, which is called mandatory liquidation in Chinese market.

\subsection{Maintenance margin and margin call}

A relative simple margin loan is considered in this paper, where one share of stock to be purchased by the loan, and the initial margin is composed of an amount of cash and a portion of the stock to be purchased. Denote the amount of cash as $Q_0$, and the proportion of the stock as $\delta$, where $Q_0 \ge 0$, and $0 \le \delta \le 1$. The initial margin ratio on the date of margin trading is
\begin{equation}\label{initial_ratio}
m_0 = \frac{Q_0 + \delta P_0}{P_0},
\end{equation}
where $P_i$ is the market price of the purchased stock on the $i$th trading day, $i=0,1,2,\cdots,T$, and $T$ is the maturity date of the loan. And the maintenance margin ratio on the date of trading is
\begin{equation}
w_0 = \frac{m_0 P_0 + P_0}{P_0}= m_0 + 1,
\end{equation}
which should be larger than or equal to the required level. Those observations give the following proposition.
\begin{proposition}\label{adequate}
Suppose the initial margin ratio on the date of margin trading  is $m_0$, then
\begin{equation*}
m_0 + 1 \ge w,
\end{equation*}
where $w$ is the required level of maintenance margin ratio. Otherwise  the initial margin is inadequate, and the customer should deposit more margin to the broker.
\end{proposition}

The cash in the initial margin can be divided into two parts on the $i$th day, one of which is the required margin $\Sigma_i$, and the other is the remaining margin $L_i$, i.e.
\begin{equation}
Q_0 \left( 1+r \right)^{i} = \Sigma_i + L_i,
\end{equation}
where $i=1,2,\cdots,T$. $\Sigma_i$ and $L_i$ satisfy the following proposition before the first margin call.
\begin{proposition}
Suppose the single day loan rate is $R$, and the riskless interest rate is $r$ per day, then the amount of required margin $\Sigma_i$ and remaining margin $L_i$ are
\begin{eqnarray}
\Sigma_i & = & w P_0 \left( 1+R \right)^{i} - (1+\delta) P_i, \\
L_i      & = & Q_0 \left( 1+r \right)^{i} -  w P_0 \left( 1+R \right)^{i} + (1+\delta) P_i.
\end{eqnarray}
\end{proposition}

\begin{remark}
When the amount of remaining margin is less than zero, a margin call should be issued. The time of margin call is determined by the amount of cash in the initial margin, the amount of the loan at time zero, the proportion of the stock in the initial margin, the two interest rates, and the dynamics of the stock prices during the period of the loan. The required initial margin ratio is not one of the key factors which determine the time of margin call.
\end{remark}

\subsection{Conditional probability of negative return}

Let $\tau$ denote  the time of first margin call, i.e.
\begin{equation}
\tau = \min \left\{ i \in \left\{1,2,\cdots,T  \right\}: L_i \le 0 \right\}.
\end{equation}
The condition of first margin call can be written as
\begin{equation}
C = \left\{
\left(1+\delta\right)P_{\tau} \le wP_0\left(1+R\right)^{\tau} -Q_{0} \left(1+r\right)^{\tau}
\right\}.
\end{equation}
It is supposed that the customer will default whenever he receives the first margin call in this paper, and the mandatory liquidation is exercised in the next trading day within the period of the loan.
Let $\tau^*$ denote the time of mandatory liquidation, which satisfies
\begin{equation}
\tau^* = \min \left\{ \tau+1, T \right\}.
\end{equation}

The return of the margin loan after mandatory liquidation is
\begin{equation}
Q_0 \left( 1+r \right)^{\tau^*} + \left( 1+\delta \right)P_{\tau^*}
   - P_0 \left( 1+R \right)^{\tau^*},
\end{equation}
and the condition of negative return is
\begin{equation}
N = \left\{
  \left( 1+\delta \right)P_{\tau^*} \le
     P_0 \left( 1+R \right)^{\tau^*}
      - Q_0 \left( 1+r \right)^{\tau^*}
 \right\}.
\end{equation}
The conditional probability of negative return (CPNR) can be written as
\begin{equation}
\text{CPNR} = \mathrm{Prob}\left\{ N \mid C\right\},
\end{equation}
where $\mathrm{Prob} \left\{ N \mid C \right\}$ is the conditional probability of $N$ given $C$.  The value of CPNR is calculated under a Markov chain model in the following sections.

\section{Markov chain and CPNR}

\subsection{Construction of Markov chain}

A Markov chain is a random process that jumps from one state to another, whose next state depends only on the present state. In order to construct a Markov model, the
daily closing stock prices are used to estimate the state space and the transition probabilities in this paper. The observed stock prices are sorted in order of increasing, and every g different prices are regarded as one state. Suppose there are n elements in the state space, denoted as $\{s_1,s_2,$ $\cdots,$ $s_n\}$, the probability of transition from the ith  to the jth state
after one step is estimated by
\begin{equation}
\hat{p}_{ij}(1)= \frac{f_{ij}}{f_{i\cdot}},
\end{equation}
where $f_{ij}$ is the observed number of transitions from the ith to the jth state  after one step, and $f_{i\cdot}=\sum_{j=1}^{n} f_{ij}$ is the total number of stock prices falling into the ith state.

The one step transition matrix can be estimated by
\begin{displaymath}
\mathbf{P(1)} = \left( \begin{array}{ccc}
\hat{p}_{11}(1) & \hat{p}_{12}(1) & \ldots \quad  \hat{p}_{1n}(1)\\
\hat{p}_{21}(1) & \hat{p}_{22}(1) & \ldots \quad  \hat{p}_{2n}(1)\\
\vdots & \vdots & \vdots\\
\hat{p}_{n1}(1) & \hat{p}_{n2}(1) & \ldots \quad  \hat{p}_{nn}(1)\\
\end{array} \right).
\end{displaymath}
Denote $\mathbf{P(n)}$ the n-step transition matrix, which can be estimated by \begin{equation}
\mathbf{P(n)}=\mathbf{P^n(1)},
\end{equation}
where the chain has been supposed to be stationary.

\subsection{Stopping time and CPNR}

The loan rate $R$ and the riskless interest rate $r$ are set to be the same for simplicity, and the condition of margin call is rewritten as
\begin{equation}
 C = \left\{  \left( 1 +\delta \right) P_{\tau} \le
    \left( wP_0 - Q_0 \right) \left( 1 + r \right)^{\tau}  \right\},
\end{equation}
where $r$ is the riskless interest rate. The condition of negative return is also rewritten as
\begin{equation}
N = \left\{ \left( 1 + \delta \right) P_{\tau^*} \le
   \left( P_0 - Q_0 \right) \left( 1+r \right)^{\tau^*} \right\}.
\end{equation}
And CPNR is rewritten as
\begin{equation}
\text{CPNR} = \mathrm{Prob} \left\{ N \mid C\right\} = \frac{\mathrm{Prob} \left\{NC\right\}}{\mathrm{Prob} \left\{ C \right\}},
\end{equation}
provided that the denominator is non zero. On the other hand, when the denominator is zero, the value of CPNR is also zero following its  definition.
In order to calculate the value of CPNR,  the numerator and denominator should be calculated
respectively.

Let
\begin{equation}
D_i= \left\{ \left( 1+\delta \right) P_{i} < \left( w P_0 - Q_0 \right) \left(  1+r \right)^{i}  \right\},
\end{equation}
where $i=1,2,\cdots,T$. The value of $\mathrm{Prob} \left\{ C \right\}$ can be calculated with the following proposition.
\begin{proposition}\label{propostion1}
Denote $C_t$ the event that broker issues the first margin call on the $t$th day, and the stock price at time zero $P_0$ is belong to the $h$th state, then
\begin{equation}
\mathrm{Prob}\left\{ C \right\} = \sum_{t=1}^{T} \mathrm{Prob} \left\{ C_t \right\},
\end{equation}
where
\begin{equation}
\mathrm{Prob}(C_t)=\mathrm{Prob}(\overline{D}_1)
\mathrm{Prob}(\overline{D}_2|\overline{D}_1)\cdots{\mathrm{Prob}(\overline{D}_{t-1}|
\overline{D}_{t-2})}\mathrm{Prob}({D_{t}}|\overline{D}_{t-1}),
\end{equation}
and
\begin{equation}
\left\{ \begin{aligned}
         \mathrm{Prob}(\overline{D}_1)&=1-\sum_{i=1}^{k_1}\hat{p}_{hi}(1),\\
         \mathrm{Prob}(D_m|\overline{D}_{m-1})&
         =\frac{\sum_{i=k_{m-1}+1}^n
                  \sum_{j=1}^{k_m}\hat{p}_{hi}(m-1)
                   \hat{p}_{ij}(1)}{\sum_{i=k_{m-1}+1}^{n}
                    \hat{p}_{hi}(m-1)},\\
  \mathrm{Prob}(\overline{D}_m|\overline{D}_{m-1}) &
         = 1-\mathrm{Prob}(D_m|\overline{D}_{m-1}),\\
         \end{aligned} \right.
                          \end{equation}
$m=2,\cdots,t$, and $k_m$ is the largest state index which satisfies
\begin{equation}
 k_m = \max \left\{ k \in \left\{ 1,2,\cdots,n \right\} : (1+\delta)q_{k} < (wP_0-Q_0)(1+r)^{m} \right\},
\end{equation}
where $q_k$ is the representative price level for the kth state.
\end{proposition}

\begin{proposition}\label{propostion2}
Denote $C_t$ the event that broker issues the first margin call on the $t$th day, and the stock price at time zero $P_0$ is belong to the $h$th state,  then the probability that the two events N and C both occur is given by
\begin{equation}
\mathrm{Prob}\left\{NC\right\}=\sum_{t=1}^T \mathrm{Prob}(NC_t),
\end{equation}
where
\begin{equation}
\begin{aligned}
         \mathrm{Prob}(NC_t)= &\mathrm{Prob}
         (\overline{D}_1)\mathrm{Prob}(\overline{D}_2|\overline{D}_1)
         \cdots \\
         &
         {\mathrm{Prob}(\overline{D}_{t-1}|
          \overline{D}_{t-2})}\mathrm{Prob}({D_{t}}|\overline{D}_{t-1})
          \mathrm{Prob}(N|D_t),
          \end{aligned}
\end{equation}
and
\begin{eqnarray}
\mathrm{Prob}(N|D_t)=
\left\{
 \begin{aligned}
           & \frac{\sum_{j=1}^{k_t}\sum_{l=1}^{a_t}\hat{p}_{hj}(t)\hat{p}_{jl}(1)}
                  {\sum_{j=1}^{k_t}\hat{p}_{hj}(t)},
                  & 1\le t \le T-1, \\
         & \frac{\sum_{j=1}^{a_{T}}\hat{p}_{hj}(T)}{\sum_{l=1}^{k_{T}}\hat{p}_{hl}(T)}, &
          t=T,
 \end{aligned}
\right.
\end{eqnarray}
where $a_{t}$ is the largest state index which satisfies
\begin{equation}
 a_{t}=\max
 \left\{
 k \in \left\{ 1,2,\cdots,n  \right\}: \left(1 + \delta \right)  q_k < (P_0-Q_0)(1+r)^{t}
 \right\},
\end{equation}
where $q_k$ is the representative price level for the kth state.
\end{proposition}

\begin{remark}
The value of CPNR can be calculated with Proposition \ref{propostion1} and \ref{propostion2} in a recursive manner. The representative price level is chosen to be the smallest one in a state.
\end{remark}

\section{Active margin system}

\subsection{Individualized maintenance margin ratio}

The required margin system is an invariable system, where the initial and maintenance margin ratios are unchangeable with respect to the dynamics of stock prices. However, the value of CPNR is determined by the current stock price $P_0$, the amount of cash  $Q_0$, the proportion of stock $\delta$, and the dynamics of stock prices during the period of loan. In order to control the risk faced by brokers below the desired level, the margin system should be different for different customers with individualized initial margins.

For a fixed value of CPNR, such as $\text{CPNR}=0.05$,  the least maintenance margin ratio
which satisfies the constraint of CPNR can be used as the required maintenance margin ratio for the customer with a particular $Q_0$ and $\delta$. Proposition \ref{adequate} can be used to determine whether the initial margin is adequate on the date of  transaction. The resulted margin system is an active system, which changes with respect to the dynamics of stock price $P_0$, the customer's individualized $Q_0$ and $\delta$. And the risk faced by brokers is controlled below the desired value of CPNR.

\subsection{Deduced margin system}

In order to give an example of margin systems deduced from the constraint of CPNR, a simple margin loan is considered in this paper, where one share of stock to be purchased using an amount of cash $Q_0$ and a portion of the same stock $\delta$ as the collateral.  The margin system can be described by the triple $\left(m,\delta,w\right)$, where $m$ is the initial margin ratio given by equation (\ref{initial_ratio}), and $w$ is the individualized maintenance margin ratio for $m$ and $\delta$.

Let
\begin{eqnarray}
 m & \in &  M= \left\{ 0.01 k, k=0,1,2,\cdots, 80 \right\},   \nonumber \\
 \delta & \in &  D= \left\{ 0.01 k, k=0,1,2,\cdots, 80 \right\}, \nonumber \\
 w & \in &  W= \left\{ 0.01 k, k=100,101,102,\cdots, 200 \right\}. \nonumber
\end{eqnarray}
The value of CPNR for a particular $\left(m,\delta,w\right)$ is denoted as $\text{CPNR}\left( m,\delta,w \right)$. For a fixed value of $\alpha$, let
\begin{equation}
\text{MDW}_{\alpha} = \left\{ \left( m,\delta,w \right) : m\in M, \delta \in D, w \in W, 1+m \ge w, \text{CPNR}\left( m,\delta,w \right) \le \alpha \right\},
\end{equation}
which is called the indifference set of margin systems. An optimal margin system $(m^*,\delta^*,w^*)$ is chosen from this set by a least squares method, which solves the following problem
\begin{eqnarray*}
& &\min_{\left( m,\delta,w \right)} \sum_{i=1}^{q}(m_i-m)^{2}+(\delta_i-\delta)^{2}+(w_i-w)^{2}, \nonumber \\
& &\text{s.t.} \left( m,\delta,w \right) \in \mathrm{MDW_{\alpha}},
\end{eqnarray*}
where $q$ is the number of elements in $\mathrm{MDW_{\alpha}}$. The resulted margin system is called the deduced margin system in this paper.

\begin{table}
\caption{Quantile analysis of the initial margin ratios under the deduced margin system.
}\label{initialtable}
\centering{
\begin{tabular*}{0.9\textwidth}{@{\extracolsep{\fill}}lllllllr}
\hline
	\textit{}&	\multicolumn{3}{c}{} & \multicolumn{4}{c}{quantiles}	 \\	
\cline{5-8}						
	\textit{Statistics}&	min	&	max	&	mean	&	0.70	&	0.80	&	 0.90	&	0.95	 \\ \hline
\textit{minimum} 	&   0.36   & 0.54    &0.46   & 0.50   & 0.52  &  0.52   & 0.52	 \\
\textit{maximum } 	&0.56    &0.76    &0.63    &0.64   & 0.64   & 0.70   & 0.71 	 \\
\textit{mean }	 &0.50 & 0.58  &0.55   & 0.56  &  0.56  &0.57  &0.57\\
\textit{quantiles}		\\	
0.20     &0.42    &0.56    &0.53    &0.54    &0.54    &0.54    &0.56\\
0.30     &0.44    &0.56    &0.54    &0.54    &0.56    &0.56    &0.56\\
0.40     &0.44    &0.58    &0.55    &0.56    &0.56    &0.56    &0.56\\
0.50     &0.46    &0.58    &0.55    &0.56    &0.56    &0.56    &0.56\\
0.60     &0.48    &0.58    &0.56    &0.56    &0.56    &0.58    &0.58\\
0.70     &0.51    &0.58    &0.57    &0.58    &0.58    &0.58    &0.58\\
0.80     &0.54    &0.61    &0.57    &0.58    &0.58    &0.59    &0.60\\
0.90     &0.56    &0.76    &0.59    &0.59    &0.60    &0.60    &0.62\\
0.95     &0.56    &0.76    &0.60    &0.60    &0.61    &0.62    &0.64\\
\hline
\end{tabular*}
}
\end{table}

\begin{table}[h]
\caption{Quantile analysis of the maintenance margin ratios under the deduced margin system.
}\label{mainteancetable}
\centering{
\begin{tabular*}{0.9\textwidth}{@{\extracolsep{\fill}}lllllllr}
\hline
		\textit{ }	&	\multicolumn{3}{c}{} & \multicolumn{4}{c}{quantiles}	 \\ \cline{5-8}						
	\textit{Statistics }&	min	&	max	&	mean	&	0.70	&	0.80	&	 0.90	&	0.95	 \\
\hline
\textit{minimum} 	&   1.02    &1.26    &1.06    &1.11    &1.15    &1.16   & 1.17	 \\
\textit{maximum } 	&1.02   & 1.45  &  1.30   & 1.35   & 1.36  &  1.38    &1.39	\\
\textit{mean }	 &1.02    &1.34  &1.18   & 1.22 & 1.24  &1.25 & 1.27\\
\textit{quantiles}	&		&		&		&		&		&		&	\\	
 0.20    &1.02    &1.32    &1.12    &1.19    &1.20    &1.22    &1.25\\
 0.30    &1.02    &1.33    &1.14    &1.21    &1.22    &1.25    &1.26\\
 0.40    &1.02    &1.35    &1.17    &1.21    &1.23    &1.25    &1.28\\
 0.50    &1.02    &1.35    &1.19    &1.23    &1.25    &1.27    &1.29\\
 0.60    &1.02    &1.35    &1.20    &1.24    &1.26    &1.29    &1.30\\
 0.70    &1.02    &1.36    &1.22    &1.27    &1.28    &1.31    &1.31\\
 0.80    &1.02    &1.37    &1.24    &1.28    &1.30    &1.32    &1.33\\
 0.90    &1.02    &1.39    &1.26    &1.29    &1.32    &1.33    &1.35\\
 0.95   &1.02    &1.40    &1.27    &1.31    &1.33    &1.35    &1.36\\
\hline
\end{tabular*}
}
\end{table}

\begin{table}[h]
\caption{Quantile analysis of the proportions of stock under the deduced margin system.}\label{proportion}
\centering{
\begin{tabular*}{0.9\textwidth}{@{\extracolsep{\fill}}lllllllr}
\hline
		&	\multicolumn{3}{c}{} & \multicolumn{4}{c}{quantiles}	 \\
					\cline{5-8}			
	\textit{Statistics}&	min	&	max	&	mean	&	0.70	&	0.80	&	 0.90	&	0.95	 \\ \hline
\textit{minimum} 	&    0.08  &  0.50     &0.39   &  0.45   &  0.45   &  0.46    & 0.46	\\
\textit{maximum } 	&0.54  &  0.75  &  0.63  &  0.67  &  0.70  &  0.71  &  0.74	\\
\textit{mean }	 &0.41   & 0.56  &  0.51  &  0.52   & 0.52  &  0.53   & 0.53\\
\textit{quantiles}	&		&		&		&		&		&		&	\\	
 0.20   &0.30 &   0.52   & 0.48  &  0.50    &0.50  &  0.50   & 0.50\\
 0.30   & 0.33   & 0.52   & 0.49   & 0.50 &   0.50  &  0.50   & 0.52\\
 0.40   &0.38 &   0.54  &  0.50 &   0.50  &  0.52  &  0.52  &  0.52\\
 0.50   &0.44  &  0.56    &0.51    &0.52  &  0.52  &  0.52   & 0.52\\
 0.60   &0.46 &   0.56  &  0.52   & 0.52  &  0.52  &  0.53  &  0.54\\
 0.70   &0.48  &  0.61   & 0.53    &0.54   & 0.54 &   0.54  &  0.54\\
 0.80   &0.50  &  0.62  &  0.54  &  0.54 &   0.54  &  0.56   & 0.56\\
 0.90   &0.52  &  0.65   & 0.56   & 0.56  &  0.58   & 0.58  &  0.60\\
 0.95  &0.52  &  0.70   & 0.57 &   0.58  &  0.59 &   0.61  &  0.63
\\
\hline
\end{tabular*}
}
\end{table}

\section{Empirical analysis of Chinese market}

\subsection{Design of out-of-sample test}

An out-of-sample test is designed in this paper to compare the performances of the required and the deduced margin systems. The value of CPNR is fixed at 0.05, and 800 price data before the current date are used to construct the Markov chain model, where every 25 different prices are regarded as one state. The period of the loans is chosen to be 30 days, and 200 loans are constructed for each stock under consideration. The sample size needed in this paper is at least 1030, and 137 stocks listed on the SSE are investigated, where the data between 26 June 2006 and 18 October 2010 are involved.

The frequency of negative returns are calculated for each stock, and those stocks whose frequencies are less than or equal to 0.05 are said to  pass the out-of-sample test. There are 126 stocks which pass the test under the deduced margin system, and the remainder can not pass the test even under the required margin system, due to the abnormal movements of stock prices. The following analysis is limited to those 126 stocks which have passed the test.

\subsection{Empirical results}

The initial and maintenance margin ratios under the deduced margin system are reported in Table \ref{initialtable} and \ref{mainteancetable}, where every statistic has 126 observations, which are calculated from the 200 observed ratios of each stock. The mean of those observed initial margin ratios is $55\%$, and the mean of those maintenance margin ratios is $118\%$.

The proportion of stock in the initial margin is
\begin{equation}
p=\frac{\delta P_0}{\delta P_0 + Q_0},
\end{equation}
which is calculated for every loan of each stock under the deduced margin system, and the results are reported in Table \ref{proportion}. The mean of those observed proportions is $51\%$.

\begin{table}
\caption{Numbers of margin calls under the required and the deduced margin systems.}\label{callnumber}
\centering{
\begin{tabular*}{\textwidth}{@{\extracolsep{\fill}}lllllllllr}
\hline
	&	\multicolumn{3}{c}{}	& \multicolumn{6}{c}{quantiles}	\\
 \cline{5-10}	
	&	min	&	max	&	mean	&	0.30 	&	0.50 	&	0.80 	&	0.90 	 &	 0.95 	&	0.99 	\\
\hline
\textit{Required}	&	5.00 &100.00  &45.79  &34.00 &46.00   &63.00 &  71.00 &77.00 &  88.00 	 \\
\textit{Deduced}	&	0.00  &  83.00  & 18.41  &2.00 & 16.00   &33.00 &  42.00 &  48.00 &  63.00 	 \\
\hline
\end{tabular*}
}
\end{table}

Frequent margin calls are regarded as one of the sources of instability in stock markets. The number of margin calls is calculated for each stock  among 200 loans  in this paper, which is the total number of loans issued at least one call during the period of  the loan. The observed results are reported in Table \ref{callnumber}.
The average number of margin calls  is  $45.79$ under the required margin system, and its counterpart under the deduced margin system is $18.41$, which is reduced by $59.8\%$ with respect to the former. The stock market under the deduced margin system is more stable than the one under the required margin system.

\begin{table}[h]
\caption{Costs of margin loans under the deduced and the required margin systems.}\label{loancost}
\centering{
\begin{tabular*}{0.9\textwidth}{@{\extracolsep{\fill}}llllllllr}
\hline
	&	\multicolumn{3}{c}{}	& \multicolumn{4}{c}{quantiles}	& \\
 \cline{5-8}
\textit{Statistics}&	min	&	max	&	mean	&	0.70 	&	0.80 	&	 0.90 	 &	0.95 	 & 	RD \\
\hline
\textit{minimum}	&	1.71   &67.94    &5.94    &6.48    &7.35    &9.34   &11.36   &0.04			 \\
	&	1.47   &62.91    &5.60   & 6.06 &   7.11   & 8.47  & 10.88	\\
\textit{maximum}	&	2.45   &97.75  & 12.24  & 13.38 &  16.18  & 20.55  & 23.50  &-0.25			 \\
	&	2.97 & 117.97   &15.19  & 17.78 &  19.79 &  24.30   &31.35	\\
\textit{mean}	&	2.18  & 83.05  &  8.76 &   9.76   &11.40   &14.36  & 17.39 & -0.04	 \\
	&	2.02   &79.12  &  8.75  & 10.00  & 11.41  & 14.71 &  18.21 	\\
\textit{quantiles}	&		&		&		&		&		&		&		&		 \\
0.20    &1.85   &73.85   & 7.31   & 8.15   & 9.48   &12.17  & 14.01  & 0.12\\
       &1.69   &67.11    &6.71    &7.33   &8.75   &11.04   &12.51\\
0.30   & 1.90   &75.57    &7.74   & 8.77   &10.40  & 12.64  & 15.75   &0.06\\
      & 1.72   &70.53    &7.31    &8.08    &9.84   &11.97   &14.80\\
0.40    &2.02  &77.25   & 8.26   & 9.34   &10.98  & 13.37  & 16.50   &0.04\\
       &1.93   &74.45    &7.87    &8.63   &10.20   &12.52  &15.82\\
0.50   & 2.12  & 86.03   & 8.83   & 9.95  & 11.52   &14.08   &17.02  &0.00\\
       &2.06   &80.18    &8.33    &9.04   &10.98   &13.93   &17.02\\
0.60    &2.18  & 88.91   & 9.24   &10.18  & 12.00  & 15.22  & 18.70  & 0.08\\
       &2.08   &82.80    &8.73    &9.58   &11.47   &14.32   &17.39\\
0.70    &2.26   &90.56   & 9.63   &10.60   &12.67  & 15.81  & 19.01  & 0.06\\
       &2.09   &84.25    &9.28   &10.14   &12.19   &15.83   &17.97\\
0.80    &2.29  & 91.70   &10.07   &11.10  & 13.25   &16.45   &19.42  &-0.09\\
       &2.12   &85.45   &10.45   &11.80   &13.86   &18.90   &21.31\\
0.90    &2.37  & 93.24  & 10.67  & 11.77  & 14.20  & 17.58  & 20.09  &-0.22\\
       &2.15   &87.21   &12.33   &14.55   &16.50   &21.53   &25.92\\
0.95   &2.41   &94.45   &11.12  & 12.12   &14.57   &18.37   &20.92  &-0.21\\
   &2.18  &111.19   &13.49  & 15.54   &17.60   &22.62   &26.61\\
\hline
\end{tabular*}
}
\end{table}

The cost of a margin loan is the terminal time value of the initial margin and the additional capital deposited to the broker to meet all of the margin calls during the period of the loan. Pursuant to the regulations of the SSE and the SZSE, the additional capital must make the maintenance margin ratio reach $150\%$, if there is a margin call. And it only need to make the maintenance margin ratio reach $w$ under the deduced margin system in this paper.
The costs under the deduced and the required margin systems are reported in Table
\ref{loancost}.  The first line of each statistic are the observations under the deduced margin system, and the second line are observed under the required margin system. And the last column of this table are the relatively difference between those two $95\%$ quantiles for the same statistic  respectively. The average cost under the deduced margin system is $8.76$, which is similar to its counterpart under the required margin system.

\section{Conclusions and discussions}

An active margin system is proposed for the margin loans using cash and stock as collateral in this paper.  Margin system can be adjusted actively according to the changes of market factors with the help of the probability of negative return (CPNR). Empirical investigations are used to compare the performances of the deduced and the required margin systems with 126 stocks listed on the SSE. It is found that the number of margin calls under the deduced margin system is significantly less than its counterpart under the required margin system, where the average costs of margin loans are similar under the two types of margin systems. Those observations indicate that CPNR can be used to measure the risk faced by brokers, and the deduced margin system   is beneficial to the stability of stock market.

The performance of individualized maintenance margin is investigated through the deduced margin system,  which is chosen from the indifference set of margin systems by a least squares method. On the other hand, margin system can be set for a particular customer with a specified initial margin. The performance of such margin system is not discussed in this paper, which will be left for further research.

% Acknowledgments here
\section*{Acknowledgments}
 {Guanghui Huang is supported by the Fundamental Research Funds for the Central Universities of China under Grant CDJZR10 100 007.}

\appendix
% the following are content of appendix

\section{Proof of proposition \ref{propostion1}}
\proof
Notice that~$C_i\cap C_j=\O$, if $i\neq j$, and $C=\sum_{t=1}^{T}C_t$,  we have
\begin{equation}
\mathrm{Prob}(C)=\sum_{t=1}^T \mathrm{Prob}(C_t).
\end{equation}
And it is directly from the definition of $C_i$ and the  Markov property of stock price, we have the following probability
\begin{equation}
\mathrm{Prob}(C_t)=\mathrm{Prob}(\overline{D}_1)
\mathrm{Prob}(\overline{D}_2|\overline{D}_1)\cdots{\mathrm{Prob}(\overline{D}_{t-1}|
\overline{D}_{t-2})}\mathrm{Prob}({D_{t}}|\overline{D}_{t-1}).
\end{equation}
Let $k_m$ denote the largest state index which satisfies
\begin{equation}
 k_m = \max \left\{ k \in \left\{ 1,2,\cdots,n \right\} : (1+\delta) q_{k} < (wP_0-Q_0)(1+r)^{m} \right\},
\end{equation}
where $q_k$ is the representative price level for the kth state. Let $s_h$ denote   the state of the current stock price,
\begin{eqnarray}
 &\mathrm{Prob}(C_1)&=\mathrm{Prob}(D_1)\nonumber \\
 & & =P\{P_1\in \left\{ s_1, s_2, \cdots, s_{k_1} \right\}|P_0\in s_h\}\nonumber \\
 & &=\sum_{i=1}^{k_1}p_{hi}(1).
\end{eqnarray}
From the Markov property of stock price, we have
\begin{eqnarray}
 & &\mathrm{Prob}\left(D_m|\overline{D}_{m-1}\right)\\
 & = &
  \frac{\mathrm{Prob}\left( \overline{D}_{m-1}D_m\right)}
   {\mathrm{Prob}\left(\overline{D}_{m-1}\right)}\nonumber \\
 & = & \frac{\mathrm{Prob}\left\{ P_m \in \left\{s_1,\cdots,s_{k_m}\right\}, P_{m-1}\in \left\{s_{k_{m-1}+1},\cdots,s_{n}\right\}|P_0 \in s_h\right\}}
 {\mathrm{Prob}\left\{P_{m-1}\in \left\{
 s_{k_{m-1}+1},s_{k_{m-1}+2},\cdots,s_n\right\}|P_0 \in s_h\right\}}\nonumber \\
 & = & \frac{\sum_{i=k_{m-1}+1}^n \sum_{j=1}^{k_m} p_{hi}(m-1)p_{ij}(1)}{\sum_{i=k_{m-1}+1}^{n}p_{hi}(m-1)},
\end{eqnarray}
which finishes the proof of Proposition \ref{propostion1}.
\endproof

\section{Proof of proposition \ref{propostion2}}

\proof
From the definition of $N$, $C$, and $C_t$, we have the following formula
\begin{equation}
 \mathrm{Prob} \left( NC \right) = \sum_{t=1}^T \mathrm{Prob}\left( NC_t\right).
\end{equation}
And from the definition of $C_t$, $D_t$ and the Markov property of stock price, we have
\begin{eqnarray}\
 \mathrm{Prob}(NC_t)&=&\mathrm{Prob}\left( N\overline{D}_1 \overline{D}_2\cdots\overline{D}_{t-1}D_t \right)\nonumber \\
  &=&\mathrm{Prob}\left(\overline{D}_1\right)
  \mathrm{Prob}\left( \overline{D}_2|\overline{D}_1\right)
  \cdots{\mathrm{Prob}\left( \overline{D}_{t-1}|\overline{D}_{t-2}\right)}\nonumber \\
  & & \mathrm{Prob}\left( {D_{t}}|\overline{D}_{t-1}\right)\mathrm{Prob}\left( N|D_{t}\right).
\end{eqnarray}
If $1 \le t \le T-1$, we have
\begin{eqnarray}
 &\mathrm{Prob}\left( N|D_{t}\right)
  & = \frac{\sum_{j=1}^{k_t}\sum_{l=1}^{a_{t}}p_{hj}(t)p_{jl}(1)}
   {\sum_{j=1}^{k_t}p_{hj}(t)},
\end{eqnarray}
where $a_t$ is the state index which satisfies
\begin{equation}
a_t = \max \left\{ k \in
       \left\{1,2,\cdots,n  \right\}: (1+\delta) q_k < \left( P_0 - Q_0 \right)(1+r)^t
          \right\}.
\end{equation}
Otherwise $i=T$, we have
\begin{eqnarray}
 &\mathrm{Prob}\left( N|D_{T}\right)
  & = \frac{\sum_{j=1}^{a_T}p_{hj}(T)}
   {\sum_{j=1}^{k_T}p_{hj}(T)},
\end{eqnarray}
which finishes the proof of Proposition \ref{propostion2}.
\endproof

\bibliographystyle{elsarticle-harv}

\begin{thebibliography}{00}

%% \bibitem must have one of the following forms:
%%   \bibitem[Jones et al.(1990)]{key}...
%%   \bibitem[Jones et al.(1990)Jones, Baker, and Williams]{key}...
%%   \bibitem[Jones et al., 1990]{key}...
%%   \bibitem[\protect\citeauthoryear{Jones, Baker, and Williams}{Jones
%%       et al.}{1990}]{key}...
%%   \bibitem[\protect\citeauthoryear{Jones et al.}{1990}]{key}...
%%   \bibitem[\protect\astroncite{Jones et al.}{1990}]{key}...
%%   \bibitem[\protect\citename{Jones et al., }1990]{key}...
%%   \harvarditem[Jones et al.]{Jones, Baker, and Williams}{1990}{key}...
%%


%1

\bibitem[Fortune(2000)]{fortune2000}
Fortune, P., 2000. Margin requirements, margin loans, and margin rates: practice and principles. New England Economic Review 9, 19-44.


%2

\bibitem[Fortune(2001)]{fortune2001}
Fortune,  P., 2001.
 Margin lending and stock market volatility.  New England Economic Review 4, 4-25.

%3

\bibitem[Ricke(2003)]{ricke2003}
Ricke, D.K.M., 2003.
 What is the link between margin loans and stock market bubbles? Dissertation, Finance Center M\"{u}nster, University of  M\"{u}nster, Germany.

 %4
 \bibitem[Galbraith(1954)]{galbraith}
Galbraith, J.K., 1954.
The great crash: 1929. Houghton Mifflin Co., New York.

%5

\bibitem[Hsieh and Miller(1990)]{hsieh1990}
Hsieh, D.A.,  Miller,   M.H., 1990.
Margin regulation and stock market volatility.  Journal of Finance
  45 (1), 3-29.


%6

\bibitem[Shiller and Bartlett(2000)]{shiller2000}
 Shiller, R., Bartlett,  B., 2000.
 Margin calls: Should the Fed step in? The Wall Street Journal, Monday, April
10, p.A46.

%7

\bibitem[Argiriou(2009)]{argiriou2009}
Argiriou, A., 2009.
 Determining margin levels using risk modelling.  Thesis,  Department of Mathematics,  Royal Institute of Technology,  Stockholm,  Sweden.


%8

\bibitem[Huang(2010)]{huang2010}
Huang, G.,  Wan,  J.,  Chen,  C., 2010.
 An active margin system and its application in Chinese margin lending market. arXiv: 1101.3974v1 [q-fin.RM].

\end{thebibliography}



\end{document}